\documentclass[aps,prc,superscriptaddress,showpacs,nofootinbib,floatfix,twocolumn]{revtex4}
\usepackage{graphicx}   % Include figure files
\usepackage{rotating}
\usepackage{dcolumn}
\usepackage{bm}
\usepackage{epsfig}
\usepackage{amssymb}
\usepackage{multirow}

\begin{document}

\title{Influence of quenched jets on di-hadron correlations}
\newcommand{\sunysb}{Department of Chemistry,
Stony Brook University, Stony Brook, NY 11794, USA}
\newcommand{\bnl}{Physics Department, Brookhaven National
Laboratory, Upton, NY 11796, USA}
\author{Jiangyong Jia}
\email[Correspond to\ ]{jjia@bnl.gov} \affiliation{\sunysb}
\affiliation{\bnl}\date{\today}
\author{Roy Lacey} \affiliation{\sunysb}

\begin{abstract}
A simple jet absorption model is used to study the influence of
hadron pairs produced by quenched jets, on di-hadron angular
correlations at intermediate transverse momentum ($p_T$). We
demonstrate that such pairs can dominate both the yield and the
shape of angular correlations, and may account for the similar
properties between the near-side ``ridge'' and the away-side
``double-humped'' structure seen in recent data. These hadron pairs
also show azimuthal anisotropy which is sensitive to the emission
angle of hadrons relative to that of the jet. Measurement of this
anisotropy may provide a constraint for elucidating the production
mechanism for near- and away-side hadron pairs at intermediate
$p_T$.
\end{abstract}

\maketitle

\section {Introduction}
A new state of nuclear matter has been created in central Au+Au
collisions ($\sqrt{s_{NN}}$= 200 GeV) at the Relativistic Heavy-ion
Collider. This matter behaves like a fluid of strongly interacting
quarks and gluons as indicated by its strong collective flow, low
kinetic shear viscosity, and large opacity~\cite{Zajc:2007ey}. It
is now commonly referred to as the strongly coupled quark-gluon
plasma (sQGP).

Energetic jets, through their absorption pattern in this medium,
can be used to probe the sQGP. Such a pattern results from the
large energy loss suffered by jets traversing the medium, which
leads to a suppression of hadron yields at high
$p_T$~\cite{Adcox:2001jp}, a phenomenon known as jet quenching. A
small fraction of jets, especially those emitted from the surface
of the interaction zone, do escape the medium. These {\it survived
jets} dominate both the single hadron and the pair yields at high
$p_T$ ($>5 $ GeV/$c$), and appear in correlation analyses as narrow
peaks at $\Delta\phi\sim0$ (near-side) and $\Delta\phi\sim\pi$
(away-side)~\cite{Adams:2006yt,Adare:2007vu}.

A majority of jets are produced inside the collision zone and are
therefore quenched. The energy and momentum of these {\it quenched
jets} are redistributed in the medium and are eventually shared
among soft hadrons. These ``medium response'' or ``feedback''
hadrons can maintain a good degree of angular correlation with the
original jets~\cite{Fries:2004hd}, depending on the details of the
energy loss mechanism and the energy dissipation process. Such
residual correlations are believed to give rise to the away-side
double-humped structure peaks at $\Delta\phi\sim\pi\pm1.1$ (the
cone)~\cite{Adler:2005ee, Adare:2008cq} and the near-side
elongation in $\Delta\eta$ (the
ridge)~\cite{Adams:2006tj,Adare:2008cq}, observed at low $p_T$ ($<4
$ GeV/$c$).

In general, correlated hadrons can be chosen from either survived
jets or quenched jets. Therefore, jet-induced pairs can be divided
into three groups,
\begin{itemize}
\item jet-jet pair: both hadrons come from fragmentation of
    survived jets.
\item jet-medium pair: one hadron comes from fragmentation of
    survived jet, the second hadron comes from medium feedback
    of quenched jet.
\item medium-medium pair: both hadrons come from medium
    feedback of quenched jets.
\end{itemize}
The three groups of pairs have very different geometric origin. For
jet-jet pairs, surface emission dominate the production of both
hadrons (e.g., tangential emission); for jet-medium pairs, surface
emission and volume emission dominate the production of the jet and
medium hadrons, respectively; for medium-medium pairs, volume
emission dominates the production of both hadrons.

In many previous correlation analyses, it is normally assumed that
one hadron (``trigger'') comes from the jet, and the second hadron
(``partner'') comes from either fragmentation or feedback, i.e.,
only jet-jet and jet-medium pairs are considered. In this picture,
the trigger comes from a jet close to the surface, which losses
some energy and fragments outside the medium. The fragments
contribute to the near-side jet peak, and the feedback of the lost
energy gives rise to the near-side ridge. Meanwhile, the away-side
jet is quenched as it traverses a longer medium, and becomes the
away-side cone.

This picture does not include the contribution from medium-medium
pairs, which do not suffer surface-bias. In a jet absorption
picture, the jet fragmentation contribution is proportional to the
number of survived jet, i.e., $\propto R^0_{AA}$ (the constant
suppression level at large $p_T$), while the medium response is
proportional to the number of quenched jets, i.e., $\propto
1-R^0_{AA}$. Therefore the yields of inter-jet pairs from jet-jet,
jet-medium, and medium-medium should scale as $(R^0_{AA})^2$,
$R^0_{AA}(1-R^0_{AA})$, and $(1-R^0_{AA})^2$, respectively. In the
limit of strong quenching, $R^{0}_{AA}\rightarrow 0$, the
medium-medium pairs dominate.

Experimentally, an enhancement of hadron pairs from quenched jets
is observed, and their spectra are found to be softer than those
from survived jets~\cite{Adare:2008cq}. Therefore, medium-medium
pairs are expected to be important only at intermediate and low
$p_T$. Indeed, the low $p_T$ auto-correlation data from the STAR
Collaboration suggests that each particle has 2--3 correlated
particles in cental Au+Au events~\cite{Daugherity:2008su}. The
modification factor for hadron-pair yields, for $p_T^A+p_T^B< 7$
GeV/$c$, has been shown to exceed the jet fragmentation
limit~\cite{Adare:2008cq}, further suggesting that feedback hadron
pairs dominate over those from jet fragmentation for
$p_T^A=p_T^B\sim 3.5$ GeV/$c$. These results imply that each jet
may induce more correlated hadrons via medium response than by
fragmentation. Furthermore, a striking similarity between the $p_T$
slopes and the particle species composition of the ridge and the
cone has been reported~\cite{Afanasiev:2007wi,
Adare:2008cq,Franz:2008ri}. The respective $p_T$ ranges over which
the yields of these hadrons are important are also the same ($<$ 4
GeV/$c$)~\cite{Adare:2008cq}. These similarities suggest that
mechanisms for the ridge and the cone could be related. They also
suggests that the bulk physics, hydrodynamical flow and parton
recombination, should play an important role in this $p_T$ region.

On the theory side, several models for the medium response to jets
have been proposed~\cite{theory}. Some involve new particle
production mechanisms such as bremsstrahlung gluon radiation or
Cerenkov gluon radiation. Others involve mechanisms related to
local heating of medium by the jet, such as momentum kick, jet
deflection, mach-cone, etc. The latter mechanisms are very
effective at generating large yield of correlated hadron pairs.
However, they have focused primarily on jet-jet and jet-medium
pairs, and because of the different geometrical origins for near-
and away-side pairs, the proposed mechanisms for the ridge and the
cone are also quite different.

An important question is the degree to which medium-medium pairs
serve to influence di-hadron correlations, and whether or not they
can give new and unifying insights on the observed medium response
for the near- and away-side jets. In this paper, we investigate the
possible role of the correlations among feedback hadrons via a
simple phenomenological jet absorption model. We do not attempt to
provide an explanation of the mechanistic origin of the jet
quenching and the medium response, but focus instead on how they
influence the relative contribution of jet-jet, jet-medium and
medium-medium pairs, and their dependence on the collision
geometry.

\section{The Model}
We start with the jet absorption picture of
Ref.~\cite{Drees:2003zh}, which assumes that jets loose all energy
after one interaction in the medium. The corresponding energy loss
probability distribution is
\begin{eqnarray}
P(\Delta E) = f\delta(\Delta E) + (1-f)\delta(\Delta E-E)
\end{eqnarray}
The survival probability $f = e^{-\kappa I}$ is controlled by
absorption strength $\kappa$ and matter integral $I$. For a parton
produced at $(x,y)$ in the transverse plane, moving along direction
$(n_x,n_y)$, the integral can be expressed as
\begin{eqnarray}
I=\int_{0}^\infty dl\hspace{2mm} l\frac{c\tau}{l+c\tau} \rho{\left(x+\left(l+c\tau\right)
n_x,y+\left(l+c\tau\right)n_y\right)}, \label{eq:2}
\end{eqnarray}
which corresponds to a quadratic dependence of absorption ($\propto
ldl$) in a longitudinal expanding medium ($\propto
c\tau/(c\tau+l)$) with a formation time of $\tau=0.2fm/c$ .

The collision geometry is modeled by a Monte Carlo Glauber
simulation of Au+Au collisions with a Woods-Saxon distribution for
the Au nucleus. We calculate the participant density profile
$\rho_{part} (x,y)$ and the binary collision density profile
$\rho_{coll} (x,y)$ in the transverse plane. Back-to-back dijet
pairs are generated according to $\rho_{coll}$ with a randomly
selected azimuthal direction. These pairs are then swum through the
medium whose density is $\rho_{part}$. We vary $\kappa$ to
reproduce the observed constant suppression factor of 4.4 for
hadrons in 0--5\% central collisions~\cite{Adler:2003au}. This
gives the value $\kappa=0.7$ that is then used for all other
centrality selections. This approach was shown to reproduce the
centrality dependence of single- and di-hadron yield at high
$p_T$~\cite{Drees:2003zh}. It decouples the details of the energy
loss process encoded in absorption strength $\kappa$, from the
geometry of the medium encoded in the integral $I$, and allows us
to focus on the geometric aspect of the interactions between jet
and medium.

The average multiplicity for a jet emitted from a given location
with a given direction in the collision zone is
\begin{eqnarray}
\langle N\rangle = \langle N_{jet}\rangle e^{-kI} +\langle N_{med}\rangle (1-e^{-kI})\quad,
\end{eqnarray}
where $N_{jet}$ is the number of fragmentation hadrons for each
survived jet and $N_{med}$ is the number of medium hadrons induced
by each quenched jet. $N_{jet}$ and $N_{med}$ are assumed to follow
Poisson distributions with a mean of 1 and 2, respectively. The
value of $\langle N_{jet}\rangle$ roughly corresponds to the number
of hadrons in 1-3 GeV/$c$ for 6 GeV/$c$ jet; the value of $\langle
N_{med}\rangle$ is chosen to account for the factor of 2-3
enhancement in the per-trigger yield seen in central Au+Au
collisions. This particular choice of $\langle N_{jet}\rangle$ and
$\langle N_{med}\rangle$ does not influence our conclusions as long
as $\langle N_{med}\rangle>\langle N_{jet}\rangle$.

For each di-jet, the total number of hadrons, $N_{tot}$, is the sum
of the multiplicity of the two jets, $N_1$ and $N_2$. $N_1$ (and
$N_2$) equals either $N_{jet}$ (if the jet survives the medium) or
$N_{med}$ (if the jet is quenched). The total number of pairs
induced by the di-jet is simply $N_{tot}(N_{tot}-1)/2$. The
azimuthal shape of these pairs is fixed by the kinematics for jet
fragmentation and the medium response as follows. The distribution
of jet hadrons is determined by the fragmentation process and
initial parton transverse momentum $k_T$. The near-side and
away-side widths are related to each other by $\sigma_{away} =
\sqrt{\sigma_{near}^2+\sigma_{k_T}^2}$, where $\sigma_{k_T}$ is the
smearing due to $k_T$. We chose $\sigma_{near}=0.2$ and
$\sigma_{k_T}=0.4$ to match the measured widths for 1-3 GeV/$c$
range in $p+p$ and $d+Au$ collisions~\cite{Adler:2005ad}. For
medium hadrons associated with a quenched jet, we distribute them
at $\pm(D=1.1)$ radian~\cite{Adare:2008cq} (relative to the initial
direction of the parton) with a Gaussian width of 0.3 radian. We
also assume that only 30\% of the inter-jet pairs are detected, so
as to account for finite detector acceptance and the swing of the
away-side jet in pseudo-rapidity.~\footnote{The typical width of
the away-side jet in $\eta$ is $\pm 2$ to $\pm3$, a detector covers
$\pm0.7$ in$\eta$ accepts about 30\% inter-jet pairs.}

We note that for models where only the jet-jet and jet-medium pairs
are included, this choice of D provides a good match for the
away-side cone structure. In our simulation, we study medium
response patterns where feedback hadrons are emitted equally on two
sides of the original jet (``mach-cone'' type) or one side of the
original jet (``deflected-jet'' type). These two patterns are
indistinguishable for jet-medium pairs, but they do lead to
different shapes for medium-medium pairs.

In di-hadron correlation analysis, one usually use per-trigger
yield (PTY) and the corresponding modification factor $I_{AA}$, the
ratio of per-trigger yield in Au+Au collisions to that in $p+p$
collision, to characterize the jet quenching and medium response.
PTY and $I_{AA}$ are good variables for high $p_T$ correlation,
since trigger yield is dominated by jet fragmentation and each
trigger tags one jet, such that per-trigger yield (PTY) is a good
representation of per-jet yield. But at $p_T<4$ GeV/$c$ region,
non-fragmentation triggers from soft production mechanisms or
medium response mechanisms become important. Experimental evidences
include a particle composition that is strongly modified relative
to that in $p+p$ collisions~\cite{Adler:2003cb}, a $R_{AA}$ and an
elliptic flow that peak in this momentum region at a level much
bigger than what one expect from the jet
quenching~\cite{Adler:2003au,Adams:2004wz}. These soft triggers
tend to dilute the $I_{AA}$, since they either have no correlation
or non-jet like correlation. This is especially important for
studying the dihadron as function of angle with respect to the
reaction plane (RP). Because the yield of soft triggers varies with
the angle {\it w.r.t} the RP, thus the dilution effect to PTY also
varies with the trigger direction.

Experimentally, such a dilution effect has been observed in a
recent PHENIX analysis~\cite{Adare:2008cq,Jia:2008dp}. As proposed
in~\cite{Adare:2008cq}, an alternative observable for describing
the medium response is the hadron pair yield (the total number of
correlated pairs per event) and the corresponding modification
factor ($J_{AA}$) in Au+Au collisions. For this reason, we present
most of the results in term of pair variables, pair yield and pair
$v_2$ (Eq.~\ref{eq:v2}).

\section{Results} 
\begin{figure}[ht]
\epsfig{file=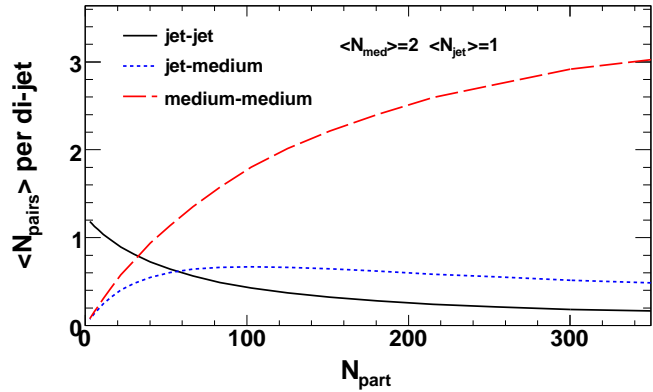,width=1.0\columnwidth}
\caption{\label{fig:1} (Color online) The centrality dependence of pair yield normalized by number of dijets
from jet-jet, jet-medium and medium-medium sources.}
\end{figure}

Figure~\ref{fig:1} compares the pair yields
for jet-jet,jet-medium and medium-medium pairs as a function of
centrality. The decrease with $N_{\rm part}$ for jet-jet pairs is
caused by the onset of jet quenching; the increase of medium-medium
pairs signifies the onset of medium response. The non-monotonic
trend for jet-medium pairs reflects the competition between jet
fragmentation and medium response. The ordering of the three
sources in central collisions, medium-medium$>$
jet-medium$>$jet-jet, is a natural consequence of $\langle
N_{med}\rangle>\langle N_{jet}\rangle$. That is, each quenched jet
gives rise to more pairs than a survived jet. It is a necessary
condition required to explain the enhancement of the pair yields.

\begin{figure}[ht]
\epsfig{file=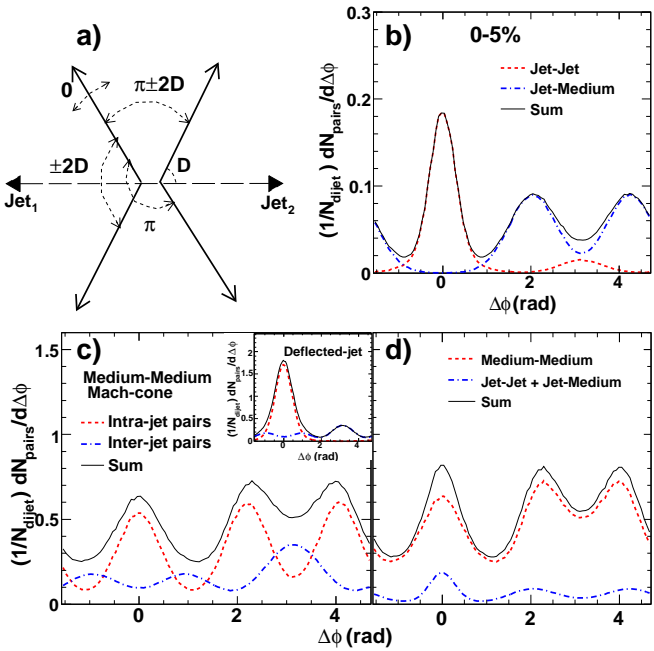,width=1\columnwidth}
\caption{\label{fig:2} (Color online) $\Delta\phi$ distributions in 0-5\% central Au+Au collisions for b) jet-jet
and jet-medium pairs, c) medium-medium pairs from mach-cone scenario (deflected-jet scenario is shown as insert), and
d) comparison of jet-jet plus jet-medium with medium-medium from mach-cone scenario, normalized by the number of generated dijets.
Panel a) illustrates the typical emission directions (solid arrows) and $\Delta\phi$ values (numbers) for medium-medium pairs.}
\end{figure}

Figure~\ref{fig:2}b shows the $\Delta\phi$ distribution for jet-jet
and jet-medium pairs, which dominates the near-side and the
away-side, respectively. There are no jet-medium pairs on the
near-side, because a single jet in our simulation does not
simultaneously produce feedback- and fragmentation-hadrons.

Figure~\ref{fig:2}c summarizes the medium-medium pairs distribution
for the mach-cone scenario. The contributions for {\it intra-jet
pairs} (referring to pairs from the same jet) and {\it inter-jet
pairs} (referring to pairs where one hadron comes from one jet and
the other comes from the opposite jet) are shown separately. The
emission directions for hadrons contributing to medium-medium pairs
are illustrated in Fig.~\ref{fig:2}a. In this picture, both jets
are quenched and converted into hadrons at angle $\pm D$ radians
relative to the original jet direction. The pairs constructed from
these hadrons, when plotted in $\Delta\phi$, should concentrate at
$\Delta\phi\sim0$, $\pi$, $\pm 2D$ and $\pi\pm2D$. The intra-jet
pairs split up into three branches at $\Delta\phi\sim0,2D$ and
$-2D$, while the inter-jet pairs have a sizable peak at
$\Delta\phi\sim\pi$ and two small satellite peaks around
$\Delta\phi\sim\pi\pm2D$. Since the value of $D=1.1$ implies $2D
\approx \pi-D$, the pairs from the same jets (at $\pm2D$) coincide
with the location of jet-medium pairs (at $\pi\pm D$).
Fig.~\ref{fig:2}d contrasts contributions from all three sources.
We see that the medium-medium pairs alone already account for the
double-humped structure on the away-side.

Figure~\ref{fig:2}d shows that the medium-medium pairs dominate the
intermediate $p_T$ correlation signal. Furthermore,
Fig.~\ref{fig:2}c shows that most of these pairs, both in the
near-side peak and in the away-side double-humps, originate from
the same jet (intra-jet pairs). Because of this common origin, the
near-side and away-side medium response pairs naturally should have
the same properties, i.e., common particle composition, common
slopes, etc. Now if there are some mechanisms which broaden the
hadrons of a jet along the pseudo-rapidity, all intra-jet
medium-medium pairs from the jet should also be elongated in
$\Delta\eta$. In this case, the experimentally observed ridge and
cone which are flat in $\Delta\eta$ simply originate from the
longitudinal broadening of those intra-jet medium-medium pairs that
appear at the near-side and the away-side,
respectively.~\footnote{Note that all intra-jet pairs, appearing on
both near- and away-side, should be elongated in $\Delta\eta$. It
is relatively easy to identify the ridge on the near-side. However,
it is harder to do so at the away-side because its net effect is
indistinguishable from the jet swing which also results in an
elongation along the beam direction.}

The insert in Fig.~\ref{fig:2}c shows the medium-medium pair
$\Delta\phi$ distribution for the deflected-jet scenario. It shows
a large peak on the near-side, but the away-side does not exhibit
the cone structure. This distribution is in stark contrast to the
ones for jet-medium pairs in Fig.~\ref{fig:2}b, which is
indistinguishable for the mach-cone and deflected-jet scenarios. In
what follows, we limit our discussion of medium-medium pairs to
those from the mach-cone scenario.

Due to the asymmetry of the overlap region of the two nuclei, the
matter integral Eq.\ref{eq:2} for a jet varies with its azimuthal
direction with respect to the reaction plane. This leads to an
azimuthal anisotropy of emitted pairs, characterized by {\it pair
$v_2$}
\begin{equation}
\label{eq:v2}
v_{2}^{\rm Pair}=\frac{\int{\cos2(\phi-\Psi_{RP}) \rm{PairYield}(\phi)d\phi}}{\int{\rm{PairYield}(\phi)d\phi}}
\end{equation}
The $\phi$ in Eq.\ref{eq:v2} can refer to either the angle of the
original jet or the angle of one of the hadron from the pair.
Motivated by earlier discussions, we use pair yield instead of
per-trigger yield for studying the reaction plane dependence,
because soft non-fragmentation triggers obscure the meaning of
per-trigger yield, and the yield of these soft triggers varies with
the angle with respect to the reaction plane.

\begin{figure}[ht]
\epsfig{file=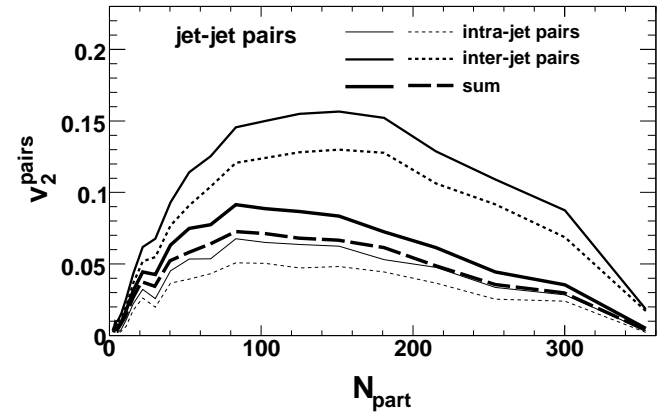,width=1.0\columnwidth}
\caption{\label{fig:3} The $v_2$ for jet-jet pairs calculated as $\langle \cos2(\phi-\Psi_{RP})\rangle$, where $\phi$ is the azimuthal angle of jets (solid lines) or fragmentation hadrons (dotted lines). }
\end{figure}

\begin{figure}[th]
\epsfig{file=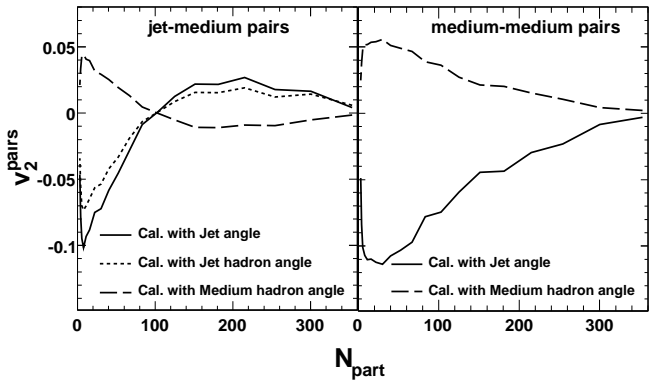,width=1\columnwidth}
\caption{\label{fig:4} The $v_2$ for jet-medium (left panel) and medium-medium (right panel) pairs
calculated as $\langle \cos2(\phi-\Psi_{RP})\rangle$, where $\phi$ is the azimuthal angle of jets (solid lines) or fragmentation hadrons (dotted lines) or medium response hadrons (dashed lines). }
\end{figure}

The $v_2$ values calculated for jet-jet, jet-medium and
medium-medium pairs are found to be different due to differences in
their geometrical origin. Fig.~\ref{fig:3} and Fig.~\ref{fig:4}
summarize the centrality dependence of the $v_2$'s for the
different pair types. The $v_2$'s for jet-jet pairs are calculated
using the $\phi$-angle of the parent jets as was done in
Ref.~\cite{Drees:2003zh} (solid lines) or by using the $\phi$-angle
of fragmentation hadrons (dotted lines). We note that the $v_2$'s
for intra-jet and
inter-jet pairs characterize the anisotropy of the survival single-
and di-jets, respectively. The former corresponds to the anisotropy
of single jet modification factor $R_{AA}(\phi-\Psi_{RP})$, i.e.
$v_2^{\rm Pair} = v_{2}^{R_{AA}}$, while the latter corresponds to
the anisotropy of di-jet modification factor $J_{AA} = R_{\rm
AA}(\phi-\Psi_{\rm RP})I_{\rm
AA}(\phi-\Psi_{RP})$~\cite{Adare:2008cq}, i.e $v_2^{\rm
Pair}\approx v_{2}^{R_{AA}}+v_{2}^{I_{AA}}$. The fact that
$v_{2}^{I_{AA}}>v_{2}^{R_{AA}}$ (Fig.~\ref{fig:3}) implies a
stronger suppression of $I_{AA}$ than $R_{AA}$ in our calculation.
The 20-30\% difference between the solid and dashed lines is due to
the smearing which results from the near- and away-side jet widths.

The $v_2$ values for the jet-medium and medium-medium pairs,
calculated using the angle of the original jets (i.e., $\phi$ in
Eq.\ref{eq:v2} refer to jet angle) are much smaller than those for
jet-jet pairs and even become negative. This is because more medium
response hadrons are produced in the out-of-plane direction as a
result of energy loss, i.e., there is more energy loss out-of-plane
than in-plane. A similar path-length dependence has been predicted
to lead to a negative $v_2$ for jet-conversion
photons~\cite{Turbide:2005bz}. However, medium response hadrons are
emitted at a large angle (D=1.1 radian) relative to the original
jet direction. Consequently, a much reduced $v_2$ is to be
expected. In fact, when the angle of these hadrons are used in our
analysis (i.e., $\phi$ in Eq.\ref{eq:v2} refer to angle of one of
the hadrons in the pair), the sign of $v_2$ is reversed, albeit
with smaller magnitudes compared to those from jet fragmentation.

The anisotropy studied in this model originates purely from
jet-quenching, other mechanisms could introduce new sources of
$v_2$ for hadron pairs, which can change the values presented in
Fig.~\ref{fig:3}. A measurement of the reaction plane dependence of
the hadron pair yield (instead of the per-trigger yield), and the
corresponding $v_2$ parameter, should be useful in helping us to
constrain the underlying mechanisms of the medium response.

In summary, we have used a simple jet absorption model to
investigate a possible influence of hadron pairs produced by
quenched jets, on di-hadron angular correlations at intermediate
$p_T$. The pair yield from quenched jets (medium-medium pair),
often neglected in model calculations, is found to be significant
when compared to jet-jet and jet-medium contributions. Using
reasonable parameters for jet kinematics, we find that the
inclusion of medium-medium pairs is sufficient to qualitatively
reproduce the shape of experimental di-hadron $\Delta\phi$
distribution. Our simulation also suggest that these pairs may
serve as a common source for the near-side ridge and away-side
cone. This finding is consistent with experimental
data~\cite{Afanasiev:2007wi, Adare:2008cq,Franz:2008ri} which show
similar properties between the ridge and the cone (i.e., similar
particle composition, spectra slope, etc). A distinct feature of
medium-medium pairs is that their yield increase with path-length
and the resulting anisotropy is very sensitive to their emission
angle relative to that of the jet (also true for jet-medium pairs
to a certain extent). The measurement of such anisotropies may
provide important constraints for the medium response to jets.

This work is supported by the NSF under award no. PHY-0701487 and
by the DOE under award no. 1011251-1-007968.

%\clearpage

\end{document}